\documentclass[preprint,journal]{vgtc}

\onlineid{0}
\preprinttext{\copyright~2026 IEEE. Author accepted manuscript (ISMAR 2026 / TVCG).}
\renewcommand{\manuscriptnotetxt}{\copyright~2026 IEEE. Personal use of this material is permitted. Permission from IEEE must be obtained for all other uses, in any current or future media, including reprinting/republishing this material for advertising or promotional purposes, creating new collective works, for resale or redistribution to servers or lists, or reuse of any copyrighted component of this work in other works.}

\vgtccategory{Research}

\title{Do We Need to ``Be There'' to Say Goodbye? Comparing Participatory Agency and Spatial Presence Across Digital Farewell Formats}

\author{
  \authororcid{Chan In SIO}{0009-0002-2701-5424},
  \authororcid{Alex MANN}{0009-0002-4248-3779},
  \authororcid{Qian Ming JIA}{0009-0000-9810-6919}, and
  \authororcid{Lik-Hang LEE*}{0000-0003-1361-1612}
}

\authorfooter{
  \item
    Chan In SIO, Alex MANN, and Qian Ming JIA are with
    The Hong Kong Polytechnic University.
    \\E-mail: \{devin.sio\,$|$\,alex.mann\,$|$\,kitty.jia\}@connect.polyu.hk.

  \item
    Lik-Hang LEE is with The Hong Kong Polytechnic University and is the
    corresponding author.
    E-mail: lik-hang.lee@polyu.edu.hk.
}

\abstract{

Digitally mediated attachment is increasingly common, yet the
mechanisms of symbolic closure in hybrid physical--digital relationships remain underexplored. We address an XR design question: how different mediated farewell formats relate to perceived symbolic closure, spatial presence, and bereavement-related emotional intensity. Using ``cloud petting'' as a controlled proxy of digitally mediated attachment, we compare three farewell formats: (1) lightweight social messaging, (2) a revisitable web memorial, and (3) immersive VR ritual participation. These formats differ in symbolic participation, spatial presence, and sensorimotor ritual enactment.
In a mixed-methods study ($N=45$), Website and VR showed
significantly higher perceived symbolic closure than Messaging, while
the present sample did not detect a significant closure advantage for
VR over Website. VR was also associated with significantly stronger
grief-related responses on the Pet Bereavement Questionnaire, whereas
negative-affect change measured by the Positive and Negative Affect
Schedule did not differ significantly across modalities.
This pattern motivates a preliminary, exploratory, mechanism-level
Agency-Amplification account: formats supporting structured farewell
participation were associated with higher perceived closure, whereas
the VR format, combining technical immersion, spatial presence, and
sensorimotor ritual enactment, was associated with stronger
loss-oriented emotional processing. Because the formats also bundled
differences in participation, revisitation, technical immersion, and
spatial presence, the study provides preliminary design guidance
without isolating the causal contribution of any single feature.

}

\keywords{Virtual Reality, Participatory Agency, Spatial Presence, Cross-Modality Comparison, Symbolic Closure, Digital Farewells}

\teaser{
  \centering
  \includegraphics[width=\linewidth, alt={Overview of the three farewell modalities and the proposed Agency-Amplification account.}]{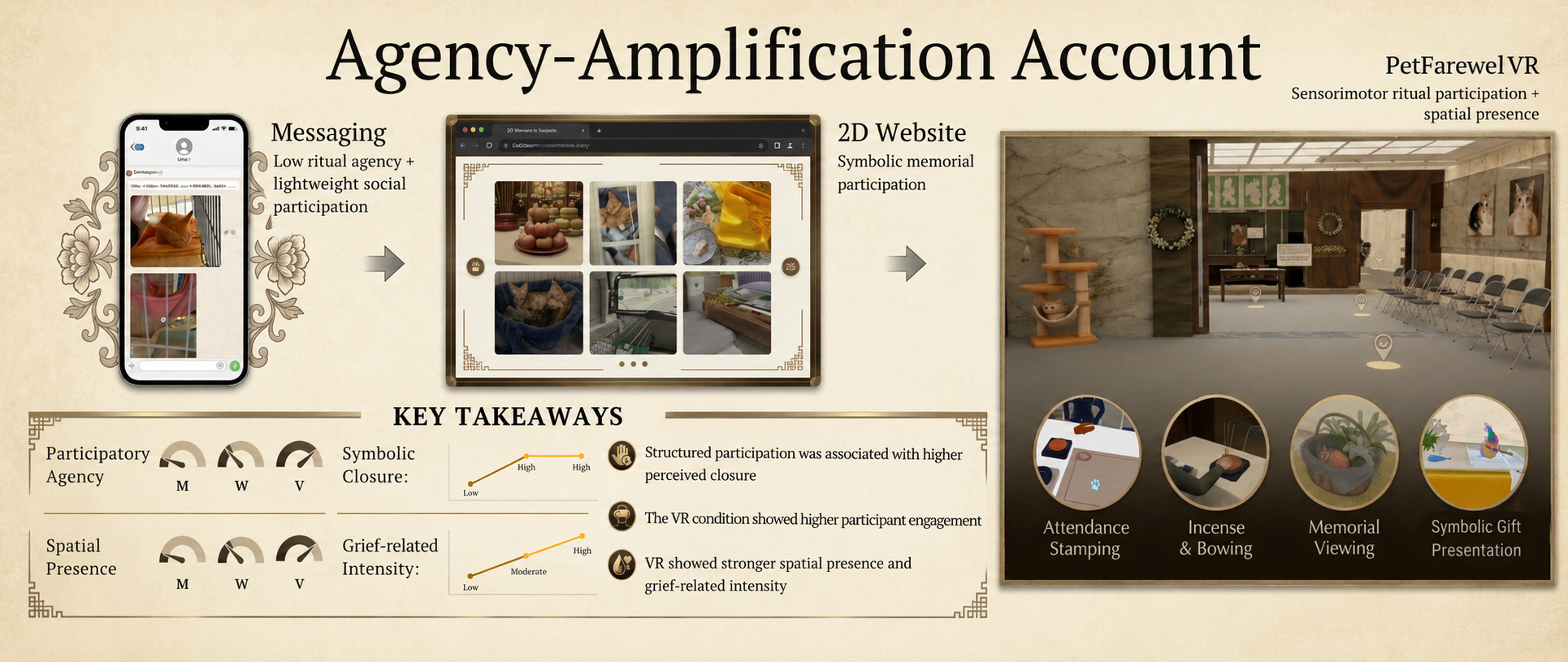}
  \par\vspace{4pt}
  {\scriptsize\sffamily\raggedright
   \refstepcounter{figure}\label{fig:teaser}
   Figure~\thefigure: Overview of the three ecologically motivated farewell formats: Messaging with low ritual agency and lightweight social participation, Website with symbolic memorial participation, and VR with sensorimotor ritual participation and spatial presence. The comparison examines how these bundled format characteristics relate to perceived symbolic closure and emotional intensity.\par}

}

\graphicspath{{figs/}{figures/}{pictures/}{images/}{./}}

\usepackage{booktabs}
\usepackage{lipsum}
\usepackage{mwe}
\usepackage{ccicons}

\usepackage{mathptmx}

\DeclareGraphicsExtensions{.pdf,.png,.jpg,.jpeg}
\begin{document}

\firstsection{Introduction}

\maketitle

Virtual Reality (VR) technologies increasingly mediate emotionally
significant experiences, including remembrance, social connection, and
loss. Ritualized farewells play an important role in acknowledging
grief, providing social support, and maintaining symbolic bonds with
the deceased
\cite{silverman1992children,Gamino2000,Burrell2020,ValeTaylor2009}.
However, the COVID-19 pandemic exposed limitations in traditional
mourning practices: restricted physical attendance and reliance on
mediated farewells were associated with psychological distress and
complicated grief responses \cite{Cardoso2020,Chen2024}.

Previous studies indicate that digital farewell formats do not support
emotional processing and symbolic closure in the same way. This raises
an XR design question: how do lightweight social messaging,
revisitable web memorial interaction, and immersive VR ritual
participation differ in perceived symbolic closure, spatial presence,
and bereavement-related emotional intensity
\cite{slater2009place,brubaker2013beyond}?

Studying these mechanisms in real-world human bereavement presents
ethical and methodological challenges. Exposing recently bereaved
individuals to experimental farewell conditions carries emotional risk,
making controlled comparison difficult. We therefore examine bonding
with non-human digital entities. In urban China, \textit{cloud petting}
describes relationships formed with pets through repeated engagement
with social-media content.\footnote{For example, platforms such as
Xiaohongshu (Little Red Book) and TikTok (Douyin in mainland China).}
Although users do not care for the animals physically, such engagement
can generate mediated presence, enhance well-being, and foster
companionship-like bonds \cite{zhou2020sense}.

Cloud-petting relationships may end abruptly when a pet disappears or
is announced deceased, often without structured acknowledgment or
ritual closure. This makes cloud petting a controlled proxy for
examining how farewell formats relate to perceived symbolic closure
within a scripted attachment scenario.
Because naturally formed cloud-petting relationships develop over
extended periods, we operationalized attachment through a scripted
10-day interaction. Although compressed and experimentally
structured, the narrative retained key features of mediated attachment
while enabling comparison across conditions.

We compared three farewell formats: (1) lightweight social messaging,
in which participants viewed materials from a physical ceremony and
could respond through text or emoji; (2) a revisitable web memorial
supporting message writing, media viewing, and virtual gifting; and
(3) immersive VR ritual participation in PetFarewellVR, combining
spatial presence with sensorimotor ritual actions such as bowing,
incense offering, and symbolic gift presentation. We examine how
these bundled formats relate to perceived symbolic closure, spatial
presence, and bereavement-related emotional intensity
\cite{slater2009place,cummings2016immersive}.
The ritual setting drew on Cantonese memorial practice to provide a
familiar cultural context for the primarily Hong Kong sample, rather
than a universal model of digital bereavement.

Our study is guided by three research questions and hypotheses
(Table~\ref{tab:rq_hypothesis_mapping}) and makes three contributions
to XR and human-computer interaction:

(1) It frames digital farewell design as an XR question concerning how
lightweight social messaging, revisitable web memorial interaction, and
immersive VR ritual participation relate to perceived symbolic closure,
spatial presence, and bereavement-related emotional intensity.

(2) It provides evidence that Website and VR, which supported
structured farewell participation, were associated with higher perceived
symbolic closure than Messaging, while VR was associated with the
strongest spatial presence and bereavement-related emotional intensity.

(3) It proposes a preliminary Agency-Amplification account and derives
design implications for calibrating participation, spatial presence, and
emotional intensity in digitally mediated farewell experiences.

\begin{table}[t]
\centering
\small
\caption{Mapping of Research Questions and Hypotheses.}
\label{tab:rq_hypothesis_mapping}
\begin{tabular}{p{0.45\linewidth} p{0.45\linewidth}}
\hline
\textbf{Research Question} & \textbf{Associated Hypothesis} \\
\hline

\textbf{RQ1:} How do differing levels of participatory agency and spatial presence across modalities relate to perceived symbolic closure in digitally mediated attachment contexts?
&
\textbf{H1:} Modalities affording structured symbolic or sensorimotor ritual participation will elicit greater perceived symbolic closure than a format with low ritual agency and lightweight social participation. \\
\hline
\textbf{RQ2:} What mechanism-level factors, including spatial presence and
sensorimotor ritual enactment, help explain modality differences in
emotional engagement and perceived closure?
&
\textbf{H2:} Greater technical immersion and sensorimotor ritual enactment will be associated with stronger spatial presence and greater emotional
engagement during the farewell experience. \\
\hline
\textbf{RQ3:} Do technical immersion and sensorimotor ritual enactment selectively amplify the emotional salience of farewell experiences while participatory symbolic action supports closure?
&
\textbf{H3:} Sensorimotor ritual enactment with spatial presence will not necessarily yield greater symbolic closure, but will be associated with stronger attachment-related grief rather than generalized negative affect.\\
\hline
\end{tabular}
\end{table}

\section{Related Work}

\subsection{Digital Mourning and Memorialization}
Research in HCI and digital media has shown that mourning practices are
not simply transferred to digital environments, but are reshaped by the
affordances of networked platforms and interactive systems
\cite{gibbs2014funeral,ulguim2018digital}. Early work in
thanatosensitive HCI argued that death, dying, and bereavement should
be treated as legitimate contexts for interaction design rather than as
marginal edge cases
\cite{massimi2009dying,massimi2011matters,albers2023dying}.
Subsequent studies demonstrated that social media and memorial
technologies can extend remembrance across time and space while
transforming how symbolic bonds with the deceased are maintained
\cite{brubaker2013beyond}. Related design research has examined
systems for legacy-making, bereavement, remembrance, and honoring
deceased loved ones \cite{gulotta2016engaging,odom2018experiences}.
Digital death research has also broadened from individual memorial
artifacts toward digital legacy, ongoingness, and end-of-life interaction
\cite{gibbs2014funeral,sas2019futures,doyle2023digital}. Studies in
Taiwan and Japan highlight the importance of culturally grounded
ritual forms in end-of-life technology design \cite{uriu2019digital}.
Recent XR and media work extends this space through telepresence for
remote attendance, virtual graveyards, AR memory sites, and responsible
AR memorialization
\cite{izumihara2019exleap,hakkila2019visiting,
pittarello2022remembering,cardenas2025lessons}.

Much of this literature emphasizes commemorative and cultural
functions more than the interaction mechanisms through which mediated
rituals may support closure. Because digital mourning experiences are
often asynchronous, screen-based, or socially distributed, it remains
difficult to determine whether relational completion arises from
symbolic participation, spatial presence, or both. This leaves open which
design features support symbolic closure when the relational context is
held constant.

\subsection{Virtual Reality for Grief and Memorialization}

Virtual reality has also been explored for grief-related support and
memorialization. Therapeutic systems such as \emph{EMMA's World}
allow users to personalize immersive environments for emotional
expression and processing \cite{Botella2008,Corno2020}. Knowles et al.
\cite{Knowles2017} showed that a real-time VR support group for
widow(er)s was feasible and acceptable compared with a grief education
website. Other studies have examined ritualized memorial interaction,
including funeral webcasting and VR tribute practices combining
symbolic action with mediated remembrance
\cite{uriu2021generating,uriu2021floral}. Research on social presence,
telepresence, and memorial-site ethics further suggests that immersive
systems can support remote participation in emotionally sensitive
remembrance contexts
\cite{wienrich2018social,izumihara2019exleap,
fisher2018ethical,mekuria2015network}.

However, prior work has generally treated VR as a standalone support
medium rather than comparing it with lower-immersion farewell formats.
It therefore remains unclear whether participatory agency supports
symbolic closure, or whether immersive interaction provides an
additional benefit.

\subsection{Presence and Emotional Amplification}

Presence and embodiment are foundational constructs in XR research.
Presence is a core mechanism through which immersive systems produce
experientially real responses \cite{slater2009place}, and meta-analytic
evidence shows that greater technological immersion increases users'
sense of presence \cite{cummings2016immersive}. Embodiment research
further suggests that immersive enactment can intensify emotional
responses. Gall et al. \cite{gall2021embodiment} found that stronger
embodiment increased emotional reactions to virtual stimuli, while
Makowski et al. \cite{makowski2017being} showed that presence
mediated the relationship between emotional arousal and memory
encoding. Thus, mechanisms that make immersive experiences feel
immediate may also amplify affective salience. Related XR work links
avatar personalization, virtual-agent emotions, and sensorimotor
interaction to presence, emotional response, and social experience
\cite{waltemate2018impact,bonsch2018social,
halbig2024common,ratcliffe2020presence}.

Prior work therefore suggests that technical immersion and
sensorimotor ritual enactment can amplify affective experience, but
their contributions to symbolic closure remain unclear. The central
question is whether spatial presence contributes to closure beyond
participatory action, or primarily amplifies the emotional salience of
the farewell. Our study addresses this gap through a
mechanism-oriented comparison of Messaging, Website, and VR within
the same scripted attachment and farewell scenario.

\section{System Design}

\subsection{Design of the Experimental Conditions}

To translate the theoretical constructs into interface features, we
distinguish four related terms. Participatory agency refers to the
ability to perform meaningful symbolic actions that contribute to the
farewell process \cite{bowman2007virtual}; it does not refer to control
over a virtual body. Technical immersion refers to system-level
affordances such as stereoscopic display, 6-DoF tracking, spatial audio,
and controller-based interaction. Spatial presence refers to the
subjective feeling of ``being there'' in a mediated environment.
Sensorimotor ritual enactment refers to bodily movement mapped to a
culturally meaningful ritual outcome, such as bowing or offering
incense. These related but distinct constructs guided the design of the
three farewell formats.

The conditions were designed as ecologically recognizable farewell
formats rather than as a factorial manipulation. Messaging afforded low
ritual agency and lightweight social participation: participants could
post text and emoji reactions and view responses from other group
members. We use \textit{lightweight social presence} to refer to this
awareness of others' participation, rather than spatial presence or
technical immersion. Website approximated a revisitable online
memorial supporting media viewing, message writing, and virtual
gifting. VR approximated attendance at a virtual funeral through
6-DoF navigation and controller-mediated ritual actions.

We intentionally retained differences in duration, setting, and access
mode. The study is therefore an ecologically motivated,
mechanism-oriented comparison of bundled affordances---including
ritual participation, revisitation, spatial presence, and sensorimotor
enactment---rather than an isolation of any single causal factor.
Messaging provided low ritual agency with lightweight social
participation, Website supported symbolic memorial participation, and
VR supported sensorimotor ritual participation with spatial presence.

PetFarewellVR did not include a visible self-avatar, and the study did
not measure body ownership or agency over a virtual body. We therefore
do not claim a full Sense of Embodiment. Where the term
\textit{embodied} is retained, it refers narrowly to bodily sensorimotor
participation.
Figure~\ref{fig:teaser} illustrates the progression from lightweight
social participation in Messaging, through screen-based symbolic
participation in Website, to sensorimotor ritual enactment in VR.

\subsection{Sensorimotor Ritual Interaction Design}

\begin{figure}
    \centering
    \includegraphics[width=\linewidth ,alt={Sensorimotor ritual interactions.}]{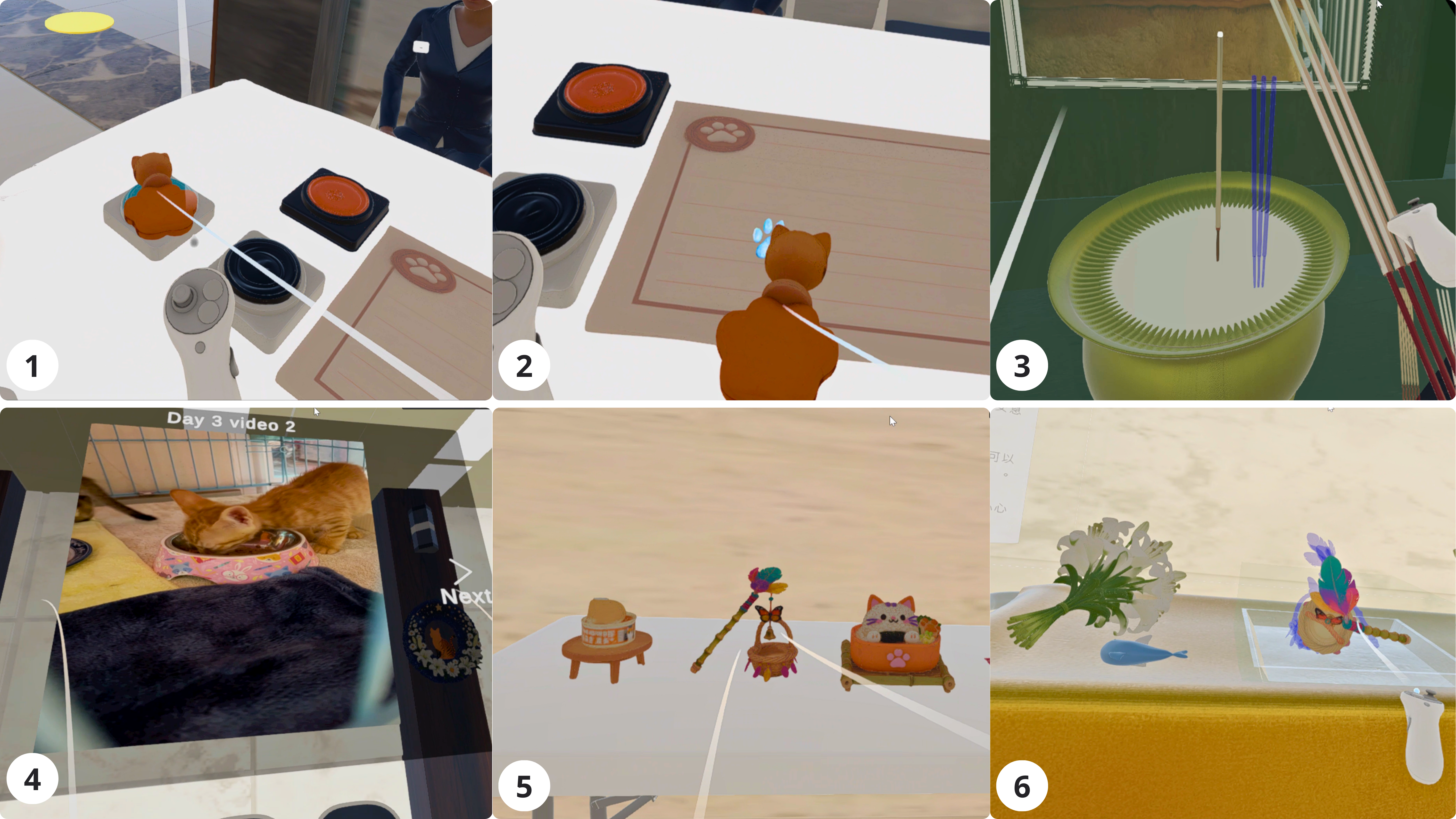}
    \caption{Sensorimotor ritual interactions: (1--2) attendance stamping, (3) incense offering and bowing, (4) memorial viewing, and (5--6) symbolic gift presentation through controller-mediated actions.}
    \label{fig:Interaction}
\end{figure}

Sensorimotor ritual enactment was operationalized through structured,
goal-directed controller actions mapped to explicit symbolic meanings.
Participants stamped their attendance, offered incense, bowed, selected
a gift, and presented it within a spatially anchored environment. These
actions directly coupled bodily movement to culturally recognizable
ritual outcomes, supporting remembrance and engagement with loss
while avoiding interactions likely to induce unrelated frustration,
anxiety, or negative affect. Figure~\ref{fig:Interaction} presents
representative stages of this sequence.

Interaction pacing followed a guided sequence based on recognizable
Cantonese funeral practices. On-screen panels briefly explained each
step and its symbolic meaning: attendance stamping signified presence,
incense expressed remembrance, bowing conveyed respect, and gift
presentation symbolized farewell and blessing. Visual cues, spatial
audio, and subtle haptic feedback reinforced action--response
contingencies and supported sensorimotor engagement.

Messaging and Website did not require bodily movement. Website
supported screen-based symbolic actions such as message posting and
virtual gifting, while Messaging supported shared messages and
reactions. Neither supported sensorimotor ritual enactment. This
comparison examined differences between screen-based symbolic
participation and VR-based sensorimotor enactment while recognizing
other format-level differences.

\subsection{Virtual Environment Design}
PetFarewellVR used a spatially coherent Cantonese-style pet funeral
hall comprising five functional areas: (1) reception, (2) an altar and
ancestral tablet area displaying the memorial portrait, (3) an incense
station, (4) an offering table, and (5) viewing and farewell rooms.
Participants progressed through attendance stamping, approaching the
memorial portrait, incense offering and bowing, gift selection,
memorial viewing, and gift presentation. Teleport anchors supported
movement, while visual instructions, spatial audio, and haptic feedback
guided each ritual action.

\begin{figure}[t]
\centering
\includegraphics[width=\linewidth,alt={topview of metagoodbye.}]{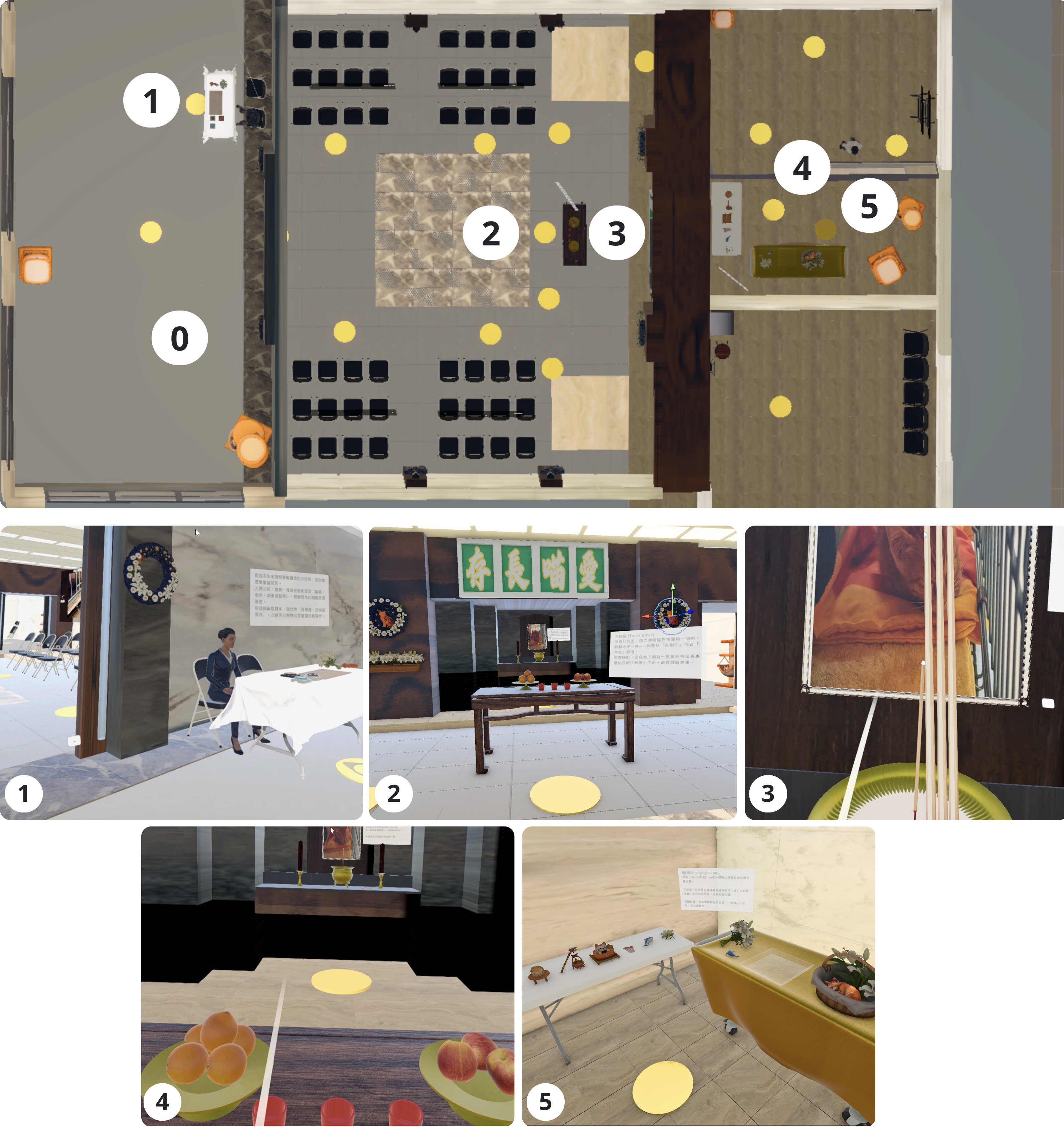}
\caption{Top view of the PetFarewellVR memorial environment and guided ritual sequence: (0) entrance, (1) reception and attendance stamping, (2) altar and ancestral tablet area displaying the memorial portrait, (3) incense station for incense offering and bowing, (4) offering table for symbolic gift selection, and (5) viewing and farewell rooms for memorial viewing and gift presentation. Yellow markers indicate teleport anchors.}
\label{fig:vr_flow_sample}
\end{figure}

The environment used recognizable Cantonese-style memorial elements
to support cultural legibility for the primarily Hong Kong sample. It
was not intended as a comprehensive or universal representation of
Cantonese funeral practice. As shown in
Figure~\ref{fig:vr_flow_sample}, the spatial sequence linked bodily
movement with symbolic ritual meaning.

\subsection{Website Memorial Design}

The Website displayed Wong Wong's memorial portrait, photographs,
and videos from the 10-day narrative. The interface did not enforce a fixed sequence: participants could review the memorial media, write a farewell message, and select a virtual symbolic gift.
The page remained accessible for 72 hours, supporting asynchronous
symbolic participation and revisitation without spatial navigation or
sensorimotor ritual enactment.

\subsection{Implementation and System Architecture}
PetFarewellVR was developed in Unity 6000.0.32f1 for standalone
deployment on Meta Quest 3. The XR Interaction Toolkit supported
6-DoF tracking and controller-based ritual interaction. Rendering was
optimized for a stable frame rate above 72 FPS, and spatial audio
provided localized feedback. Interaction events and user progression
were logged locally in JSON format.

\section{User Study}

\subsection{Study Design}

We conducted a between-subjects, ecologically motivated comparison of
three farewell formats: (1) Messaging, providing low ritual agency with
lightweight social participation; (2) Website, supporting symbolic
memorial participation; and (3) VR, supporting sensorimotor ritual
participation with spatial presence. We examined how these formats
related to perceived symbolic closure, spatial presence, and emotional
intensity.

All participants completed the same scripted 10-day cloud-petting
narrative before encountering their assigned farewell format on Day 11.
Although the pre-farewell narrative was held constant, the conditions
retained characteristic differences in duration, setting, access mode,
and interaction structure. Findings are therefore interpreted as
associations with the overall formats rather than isolated causal effects
of agency or spatial presence. To minimize risk, the study used a
scripted pet-loss scenario, which participants were informed about in
advance. The university research ethics committee approved the study
(HSEARS20250723005).

\subsection{Participants}
Forty-five participants ($N=45$) were recruited from Hong Kong
pet-lover communities through social media and volunteer networks,
with 15 participants per condition. Given the exploratory and ethically
sensitive design, the sample was intended to identify larger
between-format effects and support mixed-methods triangulation, not to
establish equivalence or rule out smaller Website--VR differences.
Eligible participants were at least 18 years old, able to read Chinese,
and reported an affinity for cats. The sample included 24 women and
21 men, most aged 25--44. VR familiarity was low to moderate
($M=2.33$, $SD=1.04$ on a 5-point scale). Individuals with recent
bereavement were excluded through ethical screening. All participants
provided informed consent and received a small voucher.

\subsection{Cloud-Petting Interaction Protocol}
Cloud petting involves remote engagement with pet updates through
comments or symbolic reactions and served here as a controlled proxy
of hybrid physical--digital attachment.

A scripted narrative was delivered through a messaging group. A
primary owner (``\textit{Janet}'') shared daily photographs, short
videos, and descriptions of a cat named \textit{Wong Wong}.
Participants acted as remote co-owners and could respond through text
or emoji, but had no physical interaction with the pet. Responses were
visible to the group, creating a shared social context. The 10-day
duration aimed to establish moderate attachment while limiting
prolonged emotional exposure.

The narrative progressed from adoption and daily interaction
(Days 1--6), through illness and deterioration (Days 7--8), to death
(Day 9) and the memorial announcement (Day 10). The farewell occurred
on Day 11, after all participants had received identical narrative
content. Table~\ref{tab:cloud_petting_summary} summarizes the
narrative; the full script and representative media are provided in the
supplementary material.

\begin{table}[t]
\centering
\small
\caption{Summary of the 10-day Cloud Petting Narrative.}
\label{tab:cloud_petting_summary}
\begin{tabular}{p{0.3cm} p{3.5cm} p{3.5cm}}
\toprule
\textbf{Day} & \textbf{Narrative Focus} & \textbf{Attachment Function} \\
\midrule
1 & Adoption announcement and introduction of Wong Wong & Initiation of relational framing \\
2 & Arrival home; shy but curious behavior & Early bonding \\
3 & Social play and adaptation to new home & Social integration \\
4 & Energetic exploration and playful attempts & Emotional engagement \\
5 & Daily routine and playful milestone & Attachment reinforcement \\
6 & Playful inconsistency; affectionate tone & Relational warmth \\
7 & Reduced appetite; signs of illness & Emotional tension \\
8 & Veterinary visit; condition worsens (morning and evening updates) & Anticipatory concern and vulnerability \\
9 & Death announcement; ``Cat Star'' narrative & Attachment disruption and grief induction \\
10 & Memorial announcement & Transition to farewell conditions \\
\bottomrule
\end{tabular}
\end{table}

\subsection{Farewell Modalities}
Participants were randomly assigned before the 10-day narrative and
encountered their assigned modality on Day 11. After completing the
farewell period, all participants received the same debriefing message
and post-study questionnaire.

\textbf{Messaging (Low Ritual Agency with Lightweight Social
Participation):}
Participants received photographs, short videos, and a textual account
of a physical farewell ceremony through the messaging group. The
materials were available for a single viewing to approximate a
time-bounded farewell update. Participants could post text and emoji
reactions and view others' responses. Messaging therefore supported
lightweight social presence and collective witnessing, but not
structured memorial actions, spatial presence, or sensorimotor ritual
participation.

\textbf{Website (Symbolic Memorial Participation):}
Participants accessed the memorial website described in Section 3.4,
where they could view media, write a farewell message, and send a
virtual symbolic gift. The page remained accessible for 72 hours,
supporting revisitation and reflection. Website outcomes therefore
reflect a bundle of symbolic participation, media engagement, and
revisitation rather than participatory agency alone.

\textbf{VR (Sensorimotor Ritual Participation with Spatial Presence):}
Participants attended the PetFarewellVR ritual, which included 6-DoF
navigation and controller-mediated actions such as attendance stamping,
incense offering, bowing, memorial viewing, and gift presentation.
Those with a compatible headset participated remotely; others attended
a supervised laboratory session using equipment provided by the
research team. Both access modes used the same instructions, ritual
sequence, and post-experience measures, although the laboratory
provided immediate technical assistance and greater environmental
control.

\section{Measures}
The measures assessed symbolic closure, grief-related response,
affective change, and spatial presence across farewell modalities
\cite{neimeyer2001meaning,collins2014interaction,
prigerson2021history,boelen2003role,watson1988development,
Thompson2007}. Farewell modality served as the experimental factor.
Spatial presence was measured as the subjective experience associated
with technical immersion, while sensorimotor ritual enactment was
implemented as a VR design feature rather than measured as a separate
embodiment construct
\cite{biocca1997cyborg,slater2009place}.

LAPS was administered before and after the 10-day attachment
narrative. Baseline PANAS was administered before the farewell. Post-farewell PANAS, PSCS-10, PBQ, IPQ, and the interviews were completed after participants had finished their assigned farewell period; for Website participants, this occurred after the 72-hour access window.

\subsection{Cloud Pet Attachment (LAPS)}
Participants completed the Lexington Attachment to Pets Scale (LAPS)
to assess attachment formed during the 10-day cloud-petting narrative
\cite{Johnson1992}. LAPS is a widely used measure of human--animal
bonding with established reliability and construct validity
\cite{Johnson1992,holcomb1985elements,zilcha2011attachment}.
It served as a manipulation check of whether the narrative established
measurable attachment before the farewell. Its 23 items were rated
from 0 (\textit{strongly disagree}) to 3 (\textit{strongly agree}), with
negatively worded items reverse-coded. The full instrument is provided
in the Supplementary Materials.

\subsection{Symbolic Closure (PSCS)}
Perceived symbolic closure refers here to the sense that a farewell
meaningfully acknowledged the relationship and provided completion,
rather than eliminated grief
\cite{field2005continuing,schut1999dual}. We developed the exploratory
10-item Perceived Symbolic Closure Scale (PSCS-10) for this
experimental context. Its four theory-informed facets were direct
closure, adequacy of ritual expression, farewell meaningfulness, and
emotional settling despite continued sadness
\cite{field2006continuing,schut1999dual}. The scale showed acceptable
internal consistency in the present sample
(Cronbach's $\alpha=.76$).

Table~\ref{tab:closure_items_structure} presents the items and
conceptual structure. Item scores were summed to produce a total from
0 to 30, with higher scores indicating greater perceived closure;
item-mean scores from 0 to 3 are also reported for interpretability.
Because several items may capture ritual completion or adequacy of
farewell enactment, and all were positively worded, the scale may be
susceptible to construct overlap and acquiescence bias. We therefore
interpret PSCS-10 as an exploratory measure of perceived symbolic
closure within this scripted context. Broader use requires further
psychometric validation.

\begin{table}[t]
\centering
\small
\caption{Perceived Symbolic Closure Scale and structure.}
\label{tab:closure_items_structure}
\begin{tabular}{p{0.24\linewidth} p{0.08\linewidth} p{0.53\linewidth}}
\hline
\textbf{Conceptual dimension} & \textbf{Item} & \textbf{Statement} \\
\hline
Direct sense of closure
& 1 & I feel that I have properly said goodbye. \\
& 2 & This farewell felt complete. \\
& 7 & I feel a sense of emotional completion after this farewell. \\

\addlinespace
Adequacy of ritual expression
& 5 & The ritual adequately expressed my feelings. \\
& 6 & I feel that the goodbye was conducted in an appropriate way. \\

\addlinespace
Meaningfulness of the farewell
& 3 & The experience gave me a sense of finishing something important. \\
& 4 & The farewell felt meaningful. \\
& 10 & I feel that this farewell marked a meaningful transition. \\

\addlinespace
Emotional settling despite sadness
& 8 & Even if I still feel sadness, the farewell feels settled. \\
& 9 & This experience helped me acknowledge the ending of the relationship. \\
\hline
\end{tabular}

\vspace{2pt}
\begin{minipage}{0.9\linewidth}
\footnotesize
\textit{Note.} Items are rated from 0 (Strongly Disagree) to 3
(Strongly Agree). No neutral midpoint was provided. All items were
positively worded and scored directly; this may increase susceptibility
to acquiescence bias.
\end{minipage}
\end{table}

\subsection{Bereavement Intensity (PBQ)}

The Pet Bereavement Questionnaire (PBQ) assessed grief-related
responses following the farewell \cite{Hunt2006}. It measures grief,
anger, guilt, and trauma and has established reliability and construct
validity in companion-animal bereavement research
\cite{Hunt2006,field2005continuing}. Items were minimally adapted to
the virtual-pet context and rated from 0 (\textit{strongly disagree}) to
3 (\textit{strongly agree}), with higher scores indicating greater
bereavement intensity. The full questionnaire is provided in the
Supplementary Materials.

\subsection{Affective State (PANAS)}
The International Short Form of the Positive and Negative Affect
Schedule (PANAS) assessed short-term affective change associated with
the farewell \cite{Thompson2007}. PANAS treats positive affect (PA)
and negative affect (NA) as largely independent dimensions, allowing
meaningful engagement to coexist with sadness or distress
\cite{watson1988development}. Participants completed it before and
after the farewell to measure change while accounting for baseline
mood. Items were rated from 1 (\textit{very slightly or not at all}) to
5 (\textit{extremely}). PA and NA were scored separately, with higher
scores indicating stronger corresponding affect. The full item list is
provided in the Supplementary Materials.

\subsection{Spatial Presence (IPQ)}
The Igroup Presence Questionnaire (IPQ) assessed the subjective sense
of ``being there,'' including spatial presence, involvement, and
experienced realism \cite{schubert2001experience}. It did not measure
Sense of Embodiment, body ownership, or agency over a virtual body
\cite{slater2009place,cummings2016immersive,kilteni2012sense}.
Items were rated from $-3$ (\textit{completely disagree}) to $+3$
(\textit{completely agree}), with higher scores indicating stronger
presence. Subscale scores followed standard IPQ procedures. The full
item list is provided in the Supplementary Materials.

\subsection{Semi-Structured Interviews}
Semi-structured interviews conducted after survey completion explored
participatory agency, spatial presence, sensorimotor ritual enactment,
symbolic closure, and emotional intensity. Questions covered five
domains aligned with RQ1--RQ3: emotional response, farewell
experience, closure and meaning, emotional intensity and regulation,
and cross-modality reflection
(Table~\ref{tab:interview_domains}).

Participants from all three modalities were interviewed to provide
balanced representation. Interviews were conducted individually,
recorded with consent, and transcribed for thematic analysis.
Transcript review and preliminary coding proceeded iteratively to
refine the thematic account across modalities.

\begin{table}[t]
\centering
\small
\setlength{\tabcolsep}{5pt}
\renewcommand{\arraystretch}{1.05}
\caption{Thematic Domains in the Interview Protocol.}
\label{tab:interview_domains}
\begin{tabular}{p{2.5cm} p{5cm}}
\toprule
\textbf{Domain} & \textbf{Focus} \\
\midrule
Emotional Response & Initial reaction to Wong Wong's death; emotional trajectory. \\
Farewell Experience & Perceived engagement, agency, authenticity in assigned modality. \\
Closure \& Meaning & Experience of symbolic completion or resolution. \\
Emotional Intensity & Appropriateness of emotional strength or amplification. \\
Cross-Modality Reflection & Comparative effectiveness and broader implications. \\
\bottomrule
\end{tabular}
\end{table}

\section{Results}

\subsection{Data Preparation and Statistical Analysis}
All questionnaire responses were screened for completeness, and no
participants were excluded for excessive missing data. LAPS, PBQ, and
PANAS composite scores were calculated by averaging the relevant item
responses. PSCS-10 items were summed to produce a total score from 0
to 30; item means from 0 to 3 are also reported for interpretability.
PANAS change scores ($\Delta$PA and $\Delta$NA) were calculated by
subtracting baseline from post-farewell scores.

One-way ANOVAs examined differences across Messaging, Website, and
VR in perceived symbolic closure (PSCS-10), bereavement intensity
(PBQ), positive-affect change, negative-affect change, and spatial
presence (IPQ). Benjamini--Hochberg false discovery rate correction
was applied across these five primary omnibus tests at $q=.05$.
Significant effects were followed by Tukey HSD comparisons. The LAPS
pre--post manipulation check was excluded from this ANOVA family.

\begin{figure}[t]
\centering

\begin{subfigure}{0.48\linewidth}
\centering
\includegraphics[width=\linewidth]{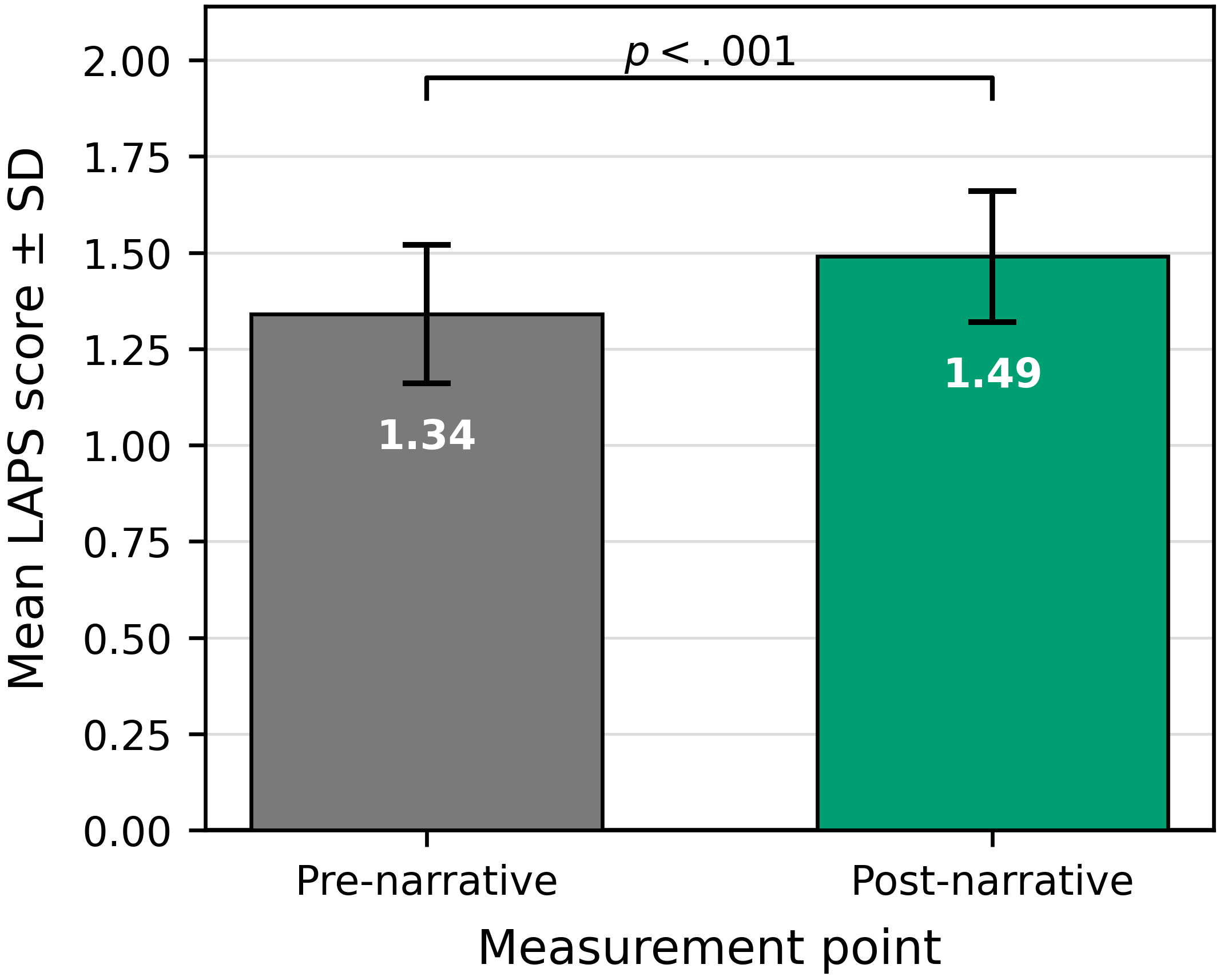}
\caption{Attachment Strength (LAPS)}
\end{subfigure}
\hfill
\begin{subfigure}{0.48\linewidth}
\centering
\includegraphics[width=\linewidth]{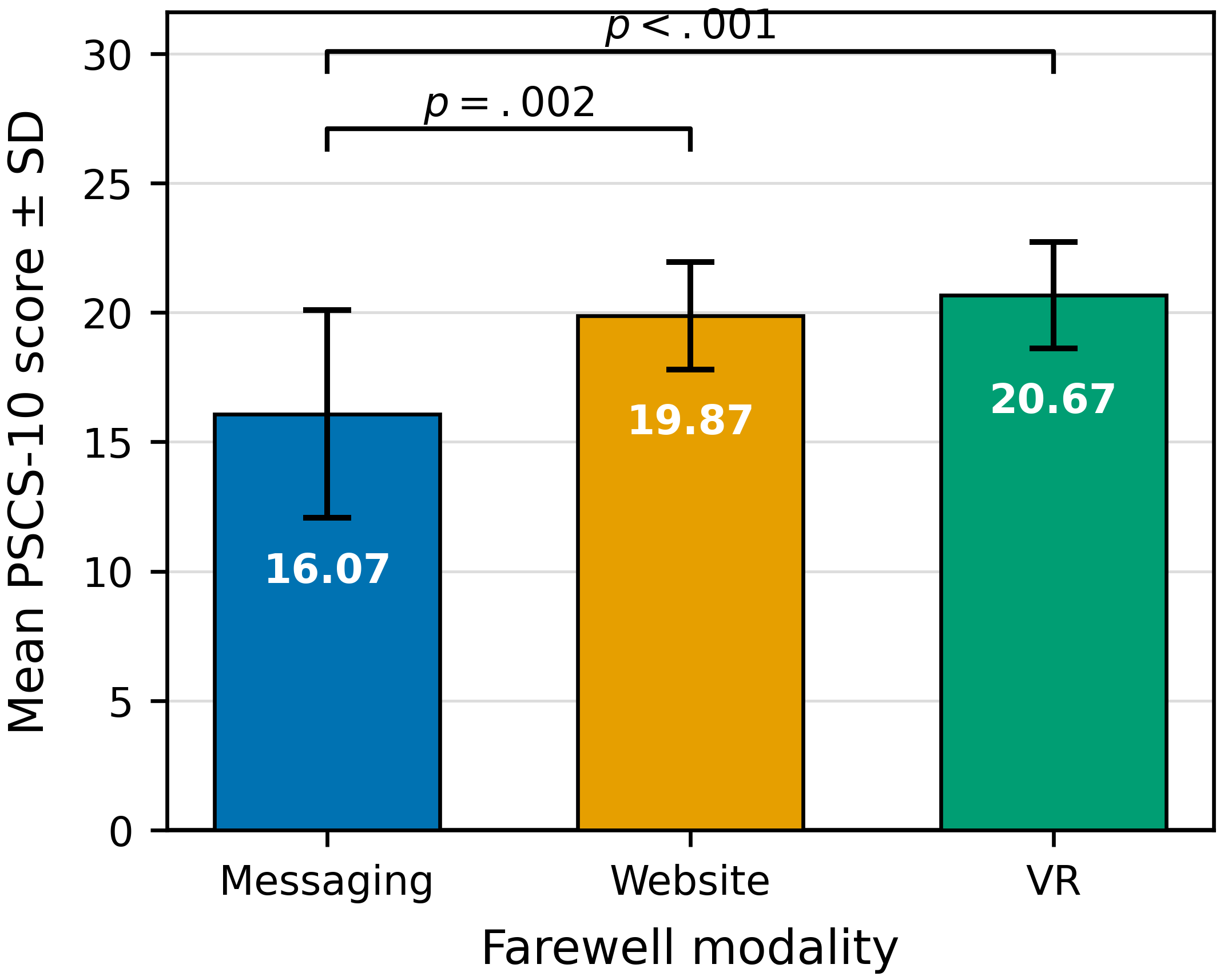}
\caption{Perceived Symbolic Closure (PSCS)}
\end{subfigure}

\vspace{0.6em}

\begin{subfigure}{0.48\linewidth}
\centering
\includegraphics[width=\linewidth]{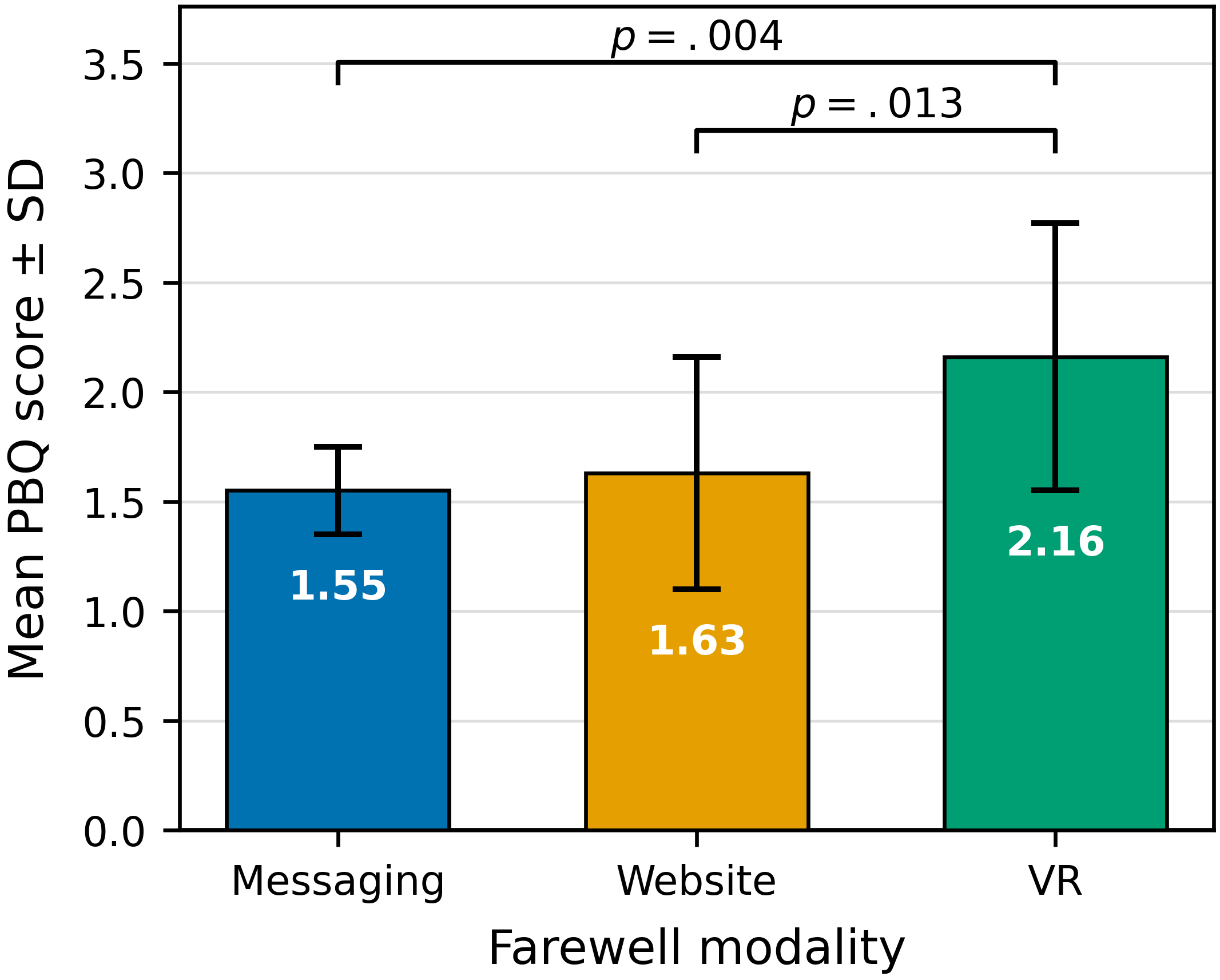}
\caption{Bereavement Intensity (PBQ)}
\end{subfigure}
\hfill
\begin{subfigure}{0.48\linewidth}
\centering
\includegraphics[width=\linewidth]{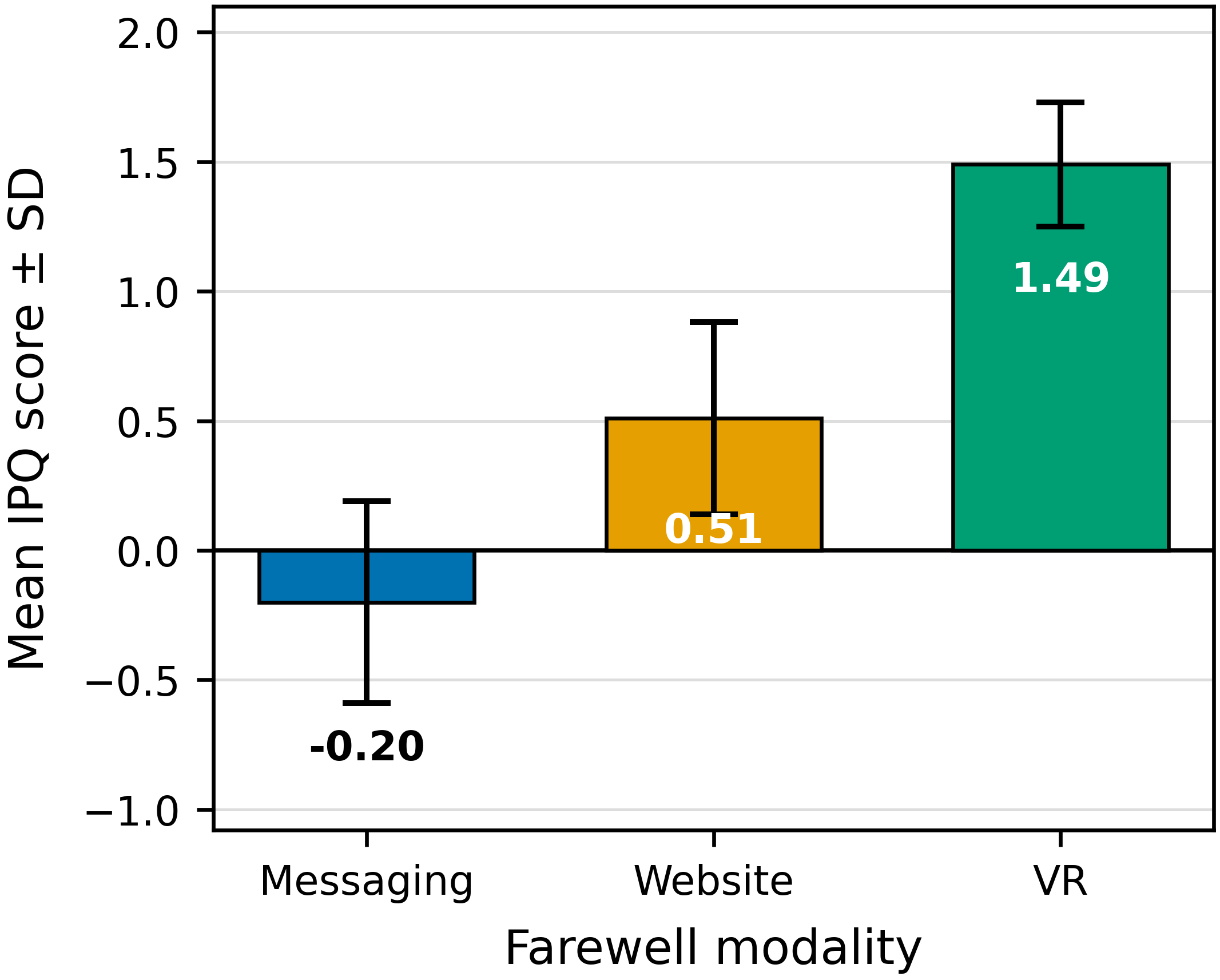}
\caption{Spatial Presence (IPQ)}
\end{subfigure}

\caption{Key quantitative outcomes across farewell modalities.
Bars indicate group means, and error bars indicate standard deviations.}
\label{fig:main_results}
\end{figure}

PSCS-10, PBQ, and IPQ remained significant after correction
(adjusted $p<.003$, $p=.005$, and adjusted $p<.003$, respectively);
positive-affect change ($p_{\mathrm{adj}}=.945$) and negative-affect
change ($p_{\mathrm{adj}}=.121$) remained non-significant. Qualitative
findings were analyzed thematically to contextualize the quantitative
results. Figure~\ref{fig:main_results} summarizes the main outcomes.

\subsection{Farewell Modality Validation: Attachment Strength}

LAPS scores increased from $M=1.34$ ($SD=0.18$) before cloud petting
to $M=1.49$ ($SD=0.17$) after the narrative. A paired-samples
$t$-test showed that this increase was significant,
$t(44)=16.52$, $p<.001$. Although attachment remained modest, the
narrative established a measurable relational baseline for the farewell
experience.

\subsection{Symbolic Closure}
Farewell modality had a significant effect on PSCS-10 scores,
$F(2,42)=11.06$, $p<.001$, $\eta^2=.35$. Scores were lowest in
Messaging ($M=16.07$, $SD=4.01$), followed by Website
($M=19.87$, $SD=2.07$) and VR ($M=20.67$, $SD=2.06$).
Tukey comparisons showed higher closure in Website than Messaging
($p=.002$) and in VR than Messaging ($p<.001$), but no significant
Website--VR difference ($p=.726$).

The item-mean scores showed the same pattern: Messaging
($M=1.61$, $SD=0.40$), Website ($M=1.99$, $SD=0.21$), and VR
($M=2.07$, $SD=0.21$). Thus, the exploratory PSCS-10 indicated
higher perceived symbolic closure in Website and VR, while the present
sample did not detect an additional closure advantage for VR over
Website.

\subsection{Bereavement Intensity}

Farewell modality significantly affected PBQ scores,
$F(2,42)=6.92$, $p=.003$. Scores were highest in VR
($M=2.16$, $SD=0.61$), followed by Website
($M=1.63$, $SD=0.53$) and Messaging
($M=1.55$, $SD=0.20$). Tukey comparisons showed higher
bereavement intensity in VR than Messaging ($p=.004$) and Website
($p=.013$), with no significant Messaging--Website difference
($p=.89$). The VR format was therefore associated with stronger
bereavement-related intensity than the two screen-based formats.

\subsection{Affective Change}

PA remained broadly stable from pre- to post-farewell
($M=3.00$ to $M=3.04$), as did NA
($M=2.56$ to $M=2.52$). As shown in
Table~\ref{tab:panas_results}, NA decreased in Messaging
($2.52\rightarrow2.32$), remained relatively stable in Website
($2.63\rightarrow2.56$), and increased in VR
($2.52\rightarrow2.68$). However, modality differences were
non-significant for both positive-affect change,
$F(2,42)=0.06$, $p=.945$, and negative-affect change,
$F(2,42)=2.46$, $p=.097$, FDR-adjusted $p=.121$.
The data therefore do not provide robust evidence of a modality effect
on short-term affective change.

\begin{table}[t]
\centering
\small
\caption{Pre- and post-farewell PANAS affect.}
\label{tab:panas_results}
\begin{tabular}{lcccccc}
\hline
 & \multicolumn{2}{c}{Pre-task} & \multicolumn{2}{c}{Post-task} & \multicolumn{2}{c}{$\Delta$} \\
Modalities & PA & NA & PA & NA & PA & NA \\
\hline
Messaging  & 2.96 & 2.52 & 3.03 & 2.32 & +0.07 & -0.20 \\
Website    & 2.99 & 2.63 & 3.00 & 2.56 & +0.01 & -0.07 \\
VR         & 3.07 & 2.52 & 3.08 & 2.68 & +0.01 & +0.16 \\
\hline
\end{tabular}
\end{table}

\subsection{Spatial Presence}
Farewell modality had a significant effect on IPQ scores,
$F(2,42)=92.94$, $p<.001$. Presence was lowest in Messaging
($M=-0.20$, $SD=0.39$), moderate in Website
($M=0.51$, $SD=0.37$), and highest in VR
($M=1.49$, $SD=0.24$). Thus, VR was associated with the strongest
subjective sense of ``being there.''

\subsection{Post-Farewell Semi-structured Interview}
Eighteen participants volunteered for interviews, six from each
modality. Transcripts were anonymized using condition-specific
identifiers (e.g., INT03-VR, INT07-WEB, INT11-MSG) and analyzed
through reflexive thematic analysis following Braun and Clarke's
six-phase framework \cite{Braun2006}. Four themes emerged:
(1) Participation as Relational Completion,
(2) Sensorimotor VR Enactment Deepening Emotional Salience,
(3) Calibrated Distance in Screen-Based Rituals, and
(4) Cultural Legibility Supporting Meaning-Making.

\textbf{Participation as Relational Completion.}
Messaging participants often described the experience as observational
(\textit{``It felt like just watching from outside''}, INT11-MSG).
Website participants identified message writing and virtual offerings
as meaningful farewell actions
(\textit{``Writing a message felt like I actually said goodbye''},
INT07-WEB). VR participants emphasized sensorimotor gestures such as
bowing and incense offering
(\textit{``When I bowed, it felt real---like I completed something''},
INT03-VR). These accounts correspond with the higher PSCS-10 scores
in Website and VR.

\textbf{Sensorimotor VR Enactment Deepening Emotional Salience.}
VR participants described the farewell as unexpectedly intense
(\textit{``It was heavier than I expected''}, INT04-VR) and emotionally
lingering
(\textit{``I still felt sad for a while''}, INT06-VR). Their accounts
focused on grief specific to the virtual pet rather than generalized
negative affect, converging with the higher PBQ scores and
non-significant PANAS results.

\textbf{Calibrated Distance in Screen-Based Rituals.}
Website participants described the experience as meaningful but
manageable. One called it \textit{``meaningful but not overwhelming''}
(INT08-WEB), while another valued being able to
\textit{``reflect quietly without feeling too emotional''}
(INT09-WEB). This theme is consistent with high closure and lower
bereavement intensity in Website than VR.

\textbf{Cultural Legibility Supporting Meaning-Making.}
Participants emphasized recognizable ritual elements. VR participants
viewed bowing and incense offering as symbolically appropriate
(\textit{``The ritual gestures made sense culturally''}, INT02-VR),
while Website participants described the memorial as resembling
familiar farewell practices
(\textit{``It felt like a proper farewell''}, INT05-WEB).

\section{Discussion}
Drawing together the quantitative and qualitative findings, this study
examined how farewell formats differing in participatory agency,
technical immersion, and sensorimotor ritual enactment related to
perceived symbolic closure, spatial presence, and
bereavement-related emotional salience. By comparing Messaging with
low ritual agency and lightweight social participation, Website with
symbolic memorial participation, and VR with sensorimotor ritual
participation and spatial presence, we examined how bundled farewell
affordances were associated with closure, presence, and emotional
intensity. The integrated findings address RQ1--RQ3 and motivate a
preliminary mechanism-level interpretation of the relationships among
participation, presence, closure, and emotional salience.

\subsection{Participatory Agency, Spatial Presence, and Symbolic Closure}

RQ1 examined how farewell formats differing in ritual agency and
spatial presence related to perceived symbolic closure. PSCS-10 scores
were higher in Website and VR than in Messaging, while the present
sample did not detect an additional closure advantage for VR over
Website. Messaging was not passive: participants could post and view
messages and emoji reactions. However, these interactions provided
lightweight social participation rather than the structured memorial
actions available in Website and VR. Website was also not a
zero-presence condition, showing moderate IPQ scores between
Messaging and VR. The observed pattern therefore cannot be attributed
to participatory agency alone.

The interviews converged with this interpretation. Under
\textit{Participation as Relational Completion}, Messaging participants
described the farewell as lacking a structured ritual action through
which to complete the goodbye, whereas Website and VR participants
identified message writing, virtual gifting, and sensorimotor ritual
actions as meaningful forms of completion. Under
\textit{Cultural Legibility Supporting Meaning-Making}, these actions
were perceived as symbolically appropriate, reinforcing relational
completion.

Prior research shows that greater immersion can increase presence and
experiential impact \cite{slater2009place,cummings2016immersive}.
However, the present results do not show that greater immersion alone
was associated with higher closure than the participatory Website
format. Website also combined symbolic participation with moderate
experiential engagement, 72-hour revisitation, and longer reflection
time. Moreover, because several PSCS-10 items concern the completeness
and adequacy of the farewell, the scores may partly reflect perceived
ritual completion. Visible Messaging responses may also have
contributed perceived support and collective witnessing.

Overall, structured farewell participation was associated with higher
perceived closure, but the study does not isolate the contributions of
agency, presence, revisitation, or social context.

\subsection{Mechanism-Level Accounts of Modality Effects}

RQ2 examined the mechanisms underlying modality differences in
closure, spatial presence, and emotional intensity. The PBQ and IPQ
results showed complementary patterns: VR showed the highest
bereavement-intensity and spatial-presence scores, whereas Website and
VR both showed higher perceived closure than Messaging. Higher PBQ
scores may reflect stronger activation of attachment-related loss
processes \cite{fraley1999loss}. This interpretation is consistent with
Continuing Bonds Theory, in which farewell practices can maintain a
symbolic connection with the lost companion
\cite{field2005continuing}, and with the Dual Process Model, in which
different formats may structure engagement with loss-oriented emotions
\cite{schut1999dual}.

Spatial presence increased progressively from Messaging through
Website to VR, indicating that the formats differed in the subjective
sense of ``being there.'' However, Website also combined symbolic
participation with moderate experiential engagement, revisitation, and
longer reflection time. The observed pattern therefore cannot be
attributed uniquely to agency or spatial presence. These complementary
patterns provide mechanism-level evidence through outcome
differentiation and qualitative convergence, rather than through a
formal mediation analysis.

\begin{table}[t]
\centering
\small
\caption{Conceptual summary of farewell-format characteristics.}
\label{tab:presence_mechanism}
\begin{tabular}{p{0.13\linewidth} p{0.1\linewidth} p{0.65\linewidth}}
\hline
\textbf{Farewell modality} & \textbf{Presence level} &
\textbf{Experiential interpretation} \\
\hline
Messaging & Low & Shared messages and emoji reactions provided
lightweight social participation and awareness of others, but the
format offered low ritual agency and no spatial interaction. \\

Website & Moderate & Message writing, media viewing, virtual gifting,
and revisitation supported symbolic memorial participation within a
screen-based format. \\

VR & High & Controller-mediated ritual actions and 6-DoF navigation
supported sensorimotor ritual participation with spatial presence. \\
\hline
\end{tabular}
\end{table}

The qualitative findings converged with this pattern. Under
\textit{Sensorimotor VR Enactment Deepening Emotional Salience}, VR
participants described the ritual as ``heavier'' and ``closer to real.''
Under \textit{Participation as Relational Completion}, Website and VR
participants linked closure to message writing, virtual gifting, and
sensorimotor ritual actions. These accounts are consistent with research
linking presence and technical immersion to affective salience
\cite{biocca1997cyborg,slater2009place,cummings2016immersive} and
sensorimotor coupling to emotional response
\cite{kilteni2012sense}.

Overall, structured farewell participation was associated with
relational completion, whereas the VR format, combining technical
immersion, spatial presence, and sensorimotor ritual enactment, was
associated with stronger bereavement-related salience. The present
study did not show that increased presence alone provided additional
closure beyond the participatory Website format.

\subsection{The Agency-Amplification Account}
RQ3 examined whether greater technical immersion, spatial presence,
and sensorimotor ritual enactment were associated with stronger
emotional salience, and whether formats supporting structured
participation were more closely associated with perceived symbolic
closure. The findings motivate a preliminary, exploratory,
mechanism-level Agency-Amplification account. This account describes
the observed pattern in which formats supporting structured farewell
participation showed higher perceived closure than Messaging, while
the VR format was associated with stronger spatial presence and
bereavement-related intensity. It should not be interpreted as a
confirmed causal model.

Both Website and VR showed higher perceived symbolic closure than
Messaging. However, the present sample did not detect an additional
closure advantage for VR over Website. This finding does not establish
equivalence between the two participatory formats and remains
compatible with smaller differences that the study was not sufficiently
sensitive to detect.

This specificity becomes clearer when comparing the two affect-related
measures. On the PANAS, changes in generalized affective states did
not differ significantly across modalities. Positive affect remained
stable, and negative-affect change did not reach statistical significance,
$F(2,42)=2.46$, $p=.097$. Although negative affect descriptively
increased in VR, the difference remained non-significant after FDR
correction. We therefore do not treat generalized negative affect as
evidence supporting the Agency-Amplification account.

By contrast, the PBQ, which more specifically captured
bereavement-related responses, showed a clear modality effect.
Participants in VR reported significantly higher bereavement intensity
than those in Messaging and Website. The VR format, which combined
technical immersion, spatial presence, and sensorimotor ritual
enactment, was therefore associated with a more specific
bereavement-related response rather than a statistically detectable
broad shift in affective state. Qualitative interviews converged with
this interpretation: VR participants more often described the experience
as ``heavier,'' emotionally lingering, and affectively absorbing.

The divergence between the PANAS and PBQ patterns helps clarify the
proposed account. The findings are more consistent with heightened
bereavement-related intensity than with a broad increase in generalized
negative affect, although the sample was not sufficiently sensitive to
rule out smaller mood-related effects. VR participants experienced the
farewell as more emotionally weighty without robust evidence of a
worsening in overall negative mood. Ritual theory also positions
enacted participation as important to symbolic completion
\cite{collins2014interaction}, which may help explain why Website
achieved high perceived closure without a statistically detectable
closure advantage for VR. Taken together, the observed pattern
associates the VR format with greater bereavement-related salience,
while formats supporting structured farewell participation were
associated with higher perceived closure.

This dissociation is summarized in Figure~\ref{fig:teaser}: structured
farewell participation was associated with higher perceived closure,
whereas the VR format was associated with greater emotional intensity
and experiential salience.

\subsection{Implications for Immersive Farewell Ritual Design}
The findings suggest that immersive ritual design should not assume a
simple linear ``more immersion is better'' relationship.

In our study, Website and VR showed numerically similar mean levels of
perceived symbolic closure. This pattern suggests that a revisitable
screen-based memorial can support perceived closure without maximal
immersion. However, Website combined symbolic participation,
moderate experiential engagement, repeated access, and longer
reflection time; the present design cannot determine the relative
contribution of these features. The results also do not establish
equivalence between Website and VR. This interpretation aligns with
XR work showing that agency, embodiment, and presence are related but
distinct design dimensions
\cite{jicol2021effects,halbig2024common}. It also resonates with
research suggesting that active interaction and user autonomy shape
affective outcomes in VR, while sensorimotor and multimodal cues may
more specifically modulate emotional salience and social meaning
\cite{cheng2024enhancing,woodworth2024design}.

Immersive ritual design should also account for the emotional intensity
associated with sensorimotor VR. Participants in VR reported higher
attachment-related bereavement intensity on the PBQ without a
corresponding statistically significant increase in generalized negative
affect on PANAS. The combination of 6-DoF interaction, spatial
presence, and sensorimotor ritual gestures was therefore associated
not merely with engagement, but with more specific loss-oriented
emotional processing in this farewell context.

Prior work on virtual memorials and relational ritual design likewise
emphasizes contribution, revisitation, and symbolic meaning-making
over sensory intensity alone
\cite{uriu2021floral,wang2023designing,uriu2025designing,
bae2025bridging}.

A foundational design priority may therefore be participatory
affordances---such as message writing, virtual gifting, and simple
offering gestures---that support relational completion. In emotionally
sensitive farewell contexts, screen-based or medium-immersion rituals
may provide a lower-intensity baseline when the goal is manageable
closure. Fully immersive VR may be used more selectively: although it
appears suited to deep and targeted affective engagement, it should be
paired with emotional scaffolding, such as pacing controls,
opportunities to pause or withdraw, clear expectations, debriefing, and
access to support.

\section{Limitations and Future Work}
The three conditions approximated recognizable farewell formats rather
than factorially isolating participatory agency, technical immersion,
spatial presence, duration, access mode, interaction structure, and
setting. Website's 72-hour access may have added revisitation and
reflection; VR's remote and laboratory deployment may have introduced
variation in technical support, comfort, environmental control,
perceived formality, and emotional safety; and visible Messaging
responses may have affected attachment, emotional response, perceived support, and
collective witnessing. Future studies should manipulate access
duration, social visibility, deployment setting, and other format
characteristics independently. The lower variability observed in Website and VR also warrants examination in future psychometric work, including assessment of possible response compression or ceiling effects.

The modest sample ($N=45$) was intended primarily to detect larger
exploratory effects and cannot establish Website--VR equivalence or
rule out smaller differences. Moreover, the scripted 10-day
cloud-petting narrative produced only moderate attachment and cannot
reproduce the depth of naturally formed bereavement. The PSCS-10 was
study-developed and exploratory; its positively worded items may be
susceptible to acquiescence bias, and its scores may partly capture
ritual completion. Larger studies should examine its factor structure,
convergent and discriminant validity, and reliability across contexts.

Generalizability is further limited by the Hong Kong sample,
Cantonese-style setting, and scripted cloud-petting scenario. The
findings should not be generalized directly to human or close-family
bereavement, other cultural contexts, or all forms of digital mourning.
Future work should examine longer-term and naturally formed
relationships, culturally diverse farewell practices, and different
forms of digitally mediated loss under appropriate ethical safeguards.

\section{Conclusion}

This study addressed an XR design question: whether immersive
``being there'' provides additional symbolic closure beyond the
structured participation available in less immersive farewell formats.
Using cloud petting as a controlled proxy of digitally mediated
attachment, we compared Messaging with low ritual agency and
lightweight social participation, Website with symbolic memorial
participation and revisitation, and VR with sensorimotor ritual
participation and spatial presence. Perceived symbolic closure was
higher in Website and VR than in Messaging, while the present sample
did not detect an additional closure advantage for VR over Website.
VR was associated with the strongest spatial presence and
bereavement-related emotional intensity, although generalized
negative-affect change did not differ significantly across formats.

These findings provide preliminary support for an exploratory,
mechanism-level Agency-Amplification account. Formats supporting
structured symbolic or sensorimotor participation were associated with
higher perceived closure, whereas the VR format, which combined
technical immersion, spatial presence, and sensorimotor ritual
enactment, was associated with greater loss-oriented emotional
salience. This pattern does not establish that participatory agency or
spatial presence independently caused the observed outcomes, because
the three farewell formats also differed in revisitation, access mode,
interaction structure, and setting.

The findings therefore do not suggest that more immersion is always
better. Meaningful participatory actions, such as message writing,
symbolic gifting, and ritual offering, may provide a foundation for
relational completion, while immersive VR may be used more
selectively when stronger emotional engagement is appropriate.
Immersive farewell experiences should be calibrated to balance
participatory meaning-making, experiential salience, and emotional
regulation. Within the scripted cloud-petting, Hong Kong, and
Cantonese-style farewell context examined here, these results provide
preliminary XR design guidance rather than a general model of human
bereavement.

\section*{Figure Credits and Copyrights}
All figures were created by the authors and remain under the authors' copyright.

\acknowledgments{

The authors acknowledge the use of AI tools for research assistance. GPT-5.2 (OpenAI) was used for grammar checking and language polishing, and Gemini 3.1 Flash Image (Google) was used to generate the illustrative images in Figure \ref{fig:teaser}.

This research is supported by the Hong Kong Polytechnic University (PolyU)'s Start-up Fund for New Recruits Under Grant (Project ID: P0046056), PolyU Department of Industrial and Systems Engineering Under Grant (Project ID: P0056354), PolyU Faculty of Engineering Under Grant (Project ID: P0064048), University of Warwick -- The Hong Kong Polytechnic University Joint Seed Fund Under Grant (Project ID: P0063938), the Research Grant Council's General Research Fund Under Grant (Project ID: P0056902), PolyU RIAM -- Research Institute for Advanced Manufacturing Under Grant (No. P0056767, P0064368), and Collaborative Project (No. P0047685, P0056286).
}

\bibliographystyle{abbrv-doi-hyperref}

\bibliography{template}

\appendix
\crefalias{section}{appendix}

\end{document}